\documentclass [12pt,a4paper,oneside] {article}

\usepackage [english] {babel}
\usepackage [T1] {fontenc}
\usepackage [latin1] {inputenc}
\usepackage{graphicx}
\usepackage{physics}
\usepackage{wasysym}
\usepackage{lipsum}
\usepackage{fancyhdr}
\usepackage{amsfonts}
\usepackage{amsmath}
\usepackage{amssymb}
\usepackage{latexsym}
\usepackage[nottoc]{tocbibind}
\usepackage{hyperref}
\usepackage{geometry}
\usepackage{eqnarray}
\usepackage{cancel}
\usepackage{verbatim}
\usepackage{bbold}
\usepackage{color}

\begin{document}

\title{Analytically Solvable Model for Qubit-Mediated Energy Transfer between Quantum Batteries}
\date{}

\author{A Crescente$^{1,2}$, Dario Ferraro $^{1,2,}$, Matteo Carrega $^{2}$ and Maura Sassetti $^{1,2}$}
\maketitle
\noindent $^{1}$ Dipartimento di Fisica, Universit\`a di Genova, Via Dodecaneso 33, 16146 Genova, Italy;\\
$^{2}$ CNR-SPIN, Via Dodecaneso 33, 16146 Genova, Italy.

\begin{abstract}
The coherent energy transfer between two identical two-level systems is investigated. Here, the first quantum system plays the role of a charger, while the second can be seen as a quantum battery. Firstly, a direct energy transfer between the two objects is considered and then compared to a transfer mediated by an additional intermediate two-level system. In this latter case, it is possible to distinguish between a two-step process, where the energy is firstly transferred from the charger to the mediator and only after from the mediator to the battery, and a single-step in which the two transfers occurs simultaneously. The differences between these configurations are discussed in the framework of an analytically solvable model completing what recently discussed in literature.   
\end{abstract}

\section{Introduction}

Over the last decades the field of quantum technologies has aroused a progressively increasing interest in the scientific community due to the possibility of manipulating and measuring miniaturized systems with very high precision~\cite{Riedel17, Acin18, Raymer19}. In this direction, the study of energy harvesting and transferring processes at the nanoscale paved the way to the development of quantum thermodynamics~\cite{Esposito09, Vinjanampathy16, Campisi16, Benenti17, Campisi17, Campaioli18, Carrega22, Cavaliere22}, where the classical laws of thermodynamics are reconsidered with the aim of describing thermal machines and energy storage devices exploiting purely quantum mechanical effects.
In particular, the concept of quantum battery (QB) was introduced by R. Alicki and M. Fannes in their seminal paper~\cite{Alicki13}. 
There, they considered the role played by entanglement in improving the storing and extraction of energy from quantum systems with respect to their classical counterparts. 
Since then, several theoretical models have been considered looking for quantum features leading to improvement of the performances in this direction~\cite{Binder15, Campaioli17, Friis18, JF20}.
Great part of the inspected models rely on platforms routinely used in the quantum computation domain, such as collections of artificial atoms~\cite{Le18, Andolina18, Rosa20, Crescente20, Carrega20, Mitchison21, Caravelli21, Seah21, Gyhm22, Crescente22, Shaghaghi23, Santos23}, and circuit quantum electrodynamics~\cite{Ferraro18, Ferraro19, Andolina19, Crescente20b, Delmonte21, Erdman22, Dou23, Gemme23}.
However, the first experimental evidence of a QB was recently presented by J. Quach {et al.} in~\cite{Quach22}, in a setup consisting of a collection of fluorescent molecules embedded in a resonant cavity. More recently, other experimental works based on quantum technology platforms have been reported. For instance, a QB based on a three-level superconducting qubit working in the transmon regime has been investigated~\cite{Hu22}. Furthermore, very promising is the opportunity to use IBM machines controlled in time to simulate the charging behaviour of QBs~\cite{Gemme22}. 

Most works in the QBs literature have been devoted to study the charging of QBs and the energy extraction from them. However, another important issue concerns the investigation of the coherent energy transfer processes that can occur between a quantum charger and a QB, allowing the realization of energetic networks able to connect distant parts of more complex quantum devices~\cite{Scarlino19, Landig19, Arrachea23}.
Preliminary theoretical studies in this direction have been proposed in \cite{Andolina18,Crescente22}, where the energy transfer processes between simple systems were analyzed. In the first work, three analytically solvable scenarios were considered, namely the direct energy transfer between a quantum harmonic oscillator (QHO) and a two-level system (TLS), or between two TLSs and finally between two QHOs. 
In the second work~\cite{Crescente22}, instead, a numerical approach was considered to understand how the presence of a mediator, e.g., a TLS or a photonic cavity, can reduce the time and improve the efficiency of the energy transfer with respect to the direct coupling between two TLSs. In passing, it is interesting to note that the study of coherent energy transfer in quantum devices presents analogies with the excitations transfer in light-harvesting photosynthetic systems~\cite{Castro08} and also with F\"orster resonant energy transfer between two light-sensitive molecules, where the first one (the donor), initially in its electronic excited state, transfers energy to the second one (the acceptor). This energy transfer is achieved  through non-radiative dipole-dipole interaction among the molecules and crucially depends on their distance and their energy mismatch~\cite{Sahoo11}.

The present work fits in the emerging field of research devoted to the study of energetic cost and management for emerging quantum  technologies~\cite{Auffeves22, Stevens22, Lewis23}. The possibility to properly control the energy transfer within quantum devices represents an important step forward in the so-called second quantum revolution~\cite{Laucht21}.
Here, we provide an analytical description of the coherent energy transfer between two identical TLSs, the first acting as a charger and the second as the QB. 
The direct energy transfer between them is compared to the case where a third identical TLS plays the role of a mediator. It is worth to underline that similar setups have been considered in the field of quantum information theory to implement high-fidelity two-qubit gates~\cite{Sung21}. In the mediated case, we identify two different energy transfer scenarios. The first being a two-step process, where the energy is initially transferred from the charger to the mediator and only after from the mediator to the QB. The second where the two transfers occurs simultaneously~\cite{Crescente22}. In all these schemes the possibility to switch on and off in a controlled way the interaction between the different parts of the system guarantees to transfer the energy only from the charger to the QB and not viceversa.
Different figures of merit are taken into account, with particular attention to the energy stored inside each building block composing the device and to the time needed to transfer the energy from the charger to the QB. As stated above, the performances will be characterized varying in time the couplings between the various parts of the total system.
 
The paper is organized as follows. In Section~\ref{models} the models for the direct and TLS-mediated cases are introduced, showing how it is possible to find analytical equations for the time evolution of the state of the total system on-resonance.
The analytical formula for the stored energy, for both the direct and the TLS-mediated scenario, are discussed in Section~\ref{Energy}. The form of the time dependent function employed to switch on and off the interaction between each part is shown in Section~\ref{FF} along with the explicit expressions for the corresponding stored energy and transfer times. The main results obtained in the direct and TLS-mediated scenarios are compared in Section~\ref{results}. Finally, Section~\ref{conclusions} is devoted to the conclusions.

\section{Model and Setting}\label{models}

The energy transfer process between two TLSs is considered. The first plays the role of a charger (C) and the second acts as a QB or as a user of the transferred energy (B). Notice that, according to the conventional literature on QB, we use the word ``charger'' to indicate a device able to provide energy to the QB~
\cite{Andolina18, Andolina19, Gemme22}. The free Hamiltonian describing this system is (hereafter we set $\hbar=1$) 
\begin{equation}
H_{0}=H_{\rm C}+H_{\rm B}
= \frac{\omega_{\rm C}}{2}\sigma_z^{\rm C}+ \frac{\omega_{\rm B}}{2}\sigma_z^{\rm B}
\end{equation}
where C and the B are characterized by energies $\omega_{\rm C}$ and $\omega_{\rm B}$ respectively and where $\sigma_z^{\rm C/B}$ are the $z$ component of Pauli matrices acting on the respective Hilbert spaces. For each TLS one can identify a ground state $|0_{\rm C/B}\rangle$ and an excited state $|1_{\rm C/B}\rangle$ such that 
\begin{eqnarray}
\sigma^{\rm {C/B}}_{z}|0_{\rm C/B}\rangle&=&-|0_{\rm C/B}\rangle\\\nonumber
\sigma^{\rm {C/B}}_{z}|1_{\rm C/B}\rangle&=&+|1_{\rm C/B}\rangle.
\end{eqnarray}
In the following the two TLSs are considered to interact both via a direct coupling between them and also with another TLS (M) which mediates their interaction. 

Notice that the following discussion will be brought up considering the total system as a closed quantum system. In general, dissipation due to interaction with the environment can affect the dynamics~\cite{Makhlin01, Weiss, Grifoni95, Calzona23}.  However, when the TLSs are weakly coupled to the environment, it is possible to accurately control and mitigate dissipation increasing the decoherence ($t_r$) and dephasing ($t_\varphi$) times~\cite{Devoret13, Wendin17}. Therefore, if one considers the dynamics of the QB for times $t \ll t_r, t_\varphi$ dissipation effects can be safely neglected~\cite{Carrega20, Caravelli21}.

\subsection{Direct Coupling Case}

\begin{figure}[h!]
\centering
\includegraphics[scale=0.4]{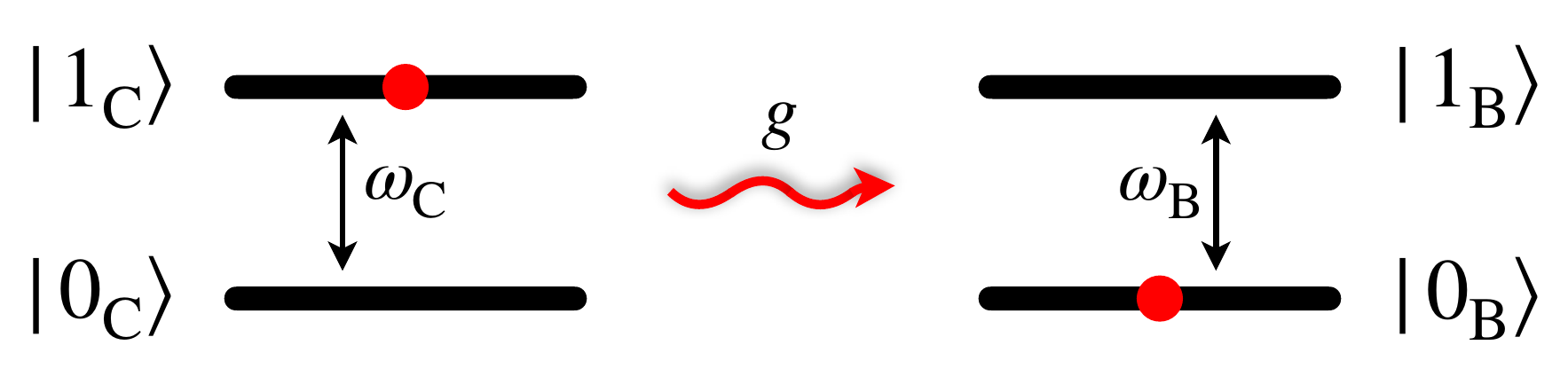} 
 \caption {Scheme of a direct energy transfer process between a quantum charger, with energy separation $\omega_{\rm C}$, and a QB, with energy separation $\omega_{\rm B}$. The coupling constant $g$ between the two TLS is modulated through a time dependent switching function.}
   \label{fig1}
\end{figure}

First of all, we consider a configuration where C and B are directly coupled (see Figure~\ref{fig1}) via a dipole-dipole interaction between the TLSs~\cite{Krantz19}. This configuration presents similarities with the F\"orster resonant energy transfer between molecules~\cite{Sahoo11}. Indeed, also in this case the dipole-dipole nature of the interaction make it relevant only for close enough quantum systems.
\vspace{-8pt}

In the rotating wave approximation (RWA)~\cite{Schweber67, Graham84, Schleich}, the complete  Hamiltonian can then be written as
\begin{equation} 
H_{\rm{dir}}(t)= H_{0}+gf(t)(\sigma_-^{\rm C}\sigma_+^{\rm B}+\sigma_+^{\rm C}\sigma_-^{\rm B}), 
\label{H_dir}
\end{equation}
where $g$ is a coupling constant, $f(t)$ is a time-dependent switching function that will be specified in the following and 
\begin{equation} \sigma^{\rm{C/B}}_\pm=\frac{\sigma^{\rm{C/B}}_x\pm i\sigma^{\rm{C/B}}_y}{2} \end{equation}
the usual ladder operators with $\sigma^{\rm{C/B}}_{x,y}$ the Pauli matrices associated to the $x$ and $y$ direction respectively. In writing the above expression, counter-rotating term of the form $\sigma_-^{\rm C}\sigma_-^{\rm B}$ and $\sigma_+^{\rm C}\sigma_+^{\rm B}$ have been safely neglected, which is a good approximation for $g\lesssim 0.1\omega_{\rm B,C}$~\cite{Schleich}. 

As the initial state of the charger-QB system, at time $t=0$, we choose 
\begin{equation}\label{insd}
|\psi(0)\rangle=|1_{\rm C}0_{\rm B}\rangle\equiv|\psi_{1}\rangle=\begin{pmatrix} 1 \\ 0 \end{pmatrix}, 
\end{equation} 
which is the tensor product of the excited state of the charger and the ground state of the~QB. 

Due to the conservation of the excitations number in the dynamics~\cite{Crescente22} it is possible to work in the basis 
\begin{eqnarray} 
|1_{\rm C}0_{\rm B}\rangle&\equiv&|\psi_{1}\rangle=\begin{pmatrix} 1 \\ 0 \end{pmatrix} \\\nonumber
|0_{\rm C}1_{\rm B}\rangle&\equiv&|\psi_{2}\rangle=\begin{pmatrix} 0 \\ 1 \end{pmatrix}, 
\end{eqnarray}
which allows to write for the chosen initial state in Equation~(\ref{insd}) the Hamiltonian in \mbox{Equation~(\ref{H_dir})} as the $2\times 2$ matrix whose elements are
\begin{equation} H_{ij}(t)=\langle \psi_i|H_{\rm{dir}}(t)|\psi_j\rangle, \end{equation}
where $i,j=1,2$. The Hamiltonian can be then written as
\begin{equation}
H_{\rm{dir}}(t)=
\begin{pmatrix}
\dfrac{\omega_{\rm C}-\omega_{\rm B}}{2} & gf(t) \\\\
gf(t) & -\dfrac{\omega_{\rm C}-\omega_{\rm B}}{2}
\end{pmatrix}.
\end{equation}
To simplify the model we choose to work with identical TLSs ($\omega_{\rm C}=\omega_{\rm B}$), for which one~has
\begin{equation}H_{\rm{dir}}(t)=
\begin{pmatrix}
0 & gf(t) \\\\
gf(t) & 0
\end{pmatrix}.
\end{equation}

%%%%%%%%%%%%%%%%%%%%%%%%%%%%%

To study the dynamics of the system it is necessary to solve the Schr\"odinger equation 
\begin{equation}\label{SE} i\frac{d}{dt}|\psi(t)\rangle=H_{\rm{dir}}(t)|\psi(t)\rangle, \end{equation}
where the state evolved in time $|\psi(t)\rangle$ can be written in spinorial form as
\begin{equation}
|\psi(t)\rangle=a_{1}(t)|\psi_{1}\rangle+a_{2}(t)|\psi_{2}\rangle \equiv
\begin{pmatrix}
a_1(t) \\
a_2(t)
\end{pmatrix}.
\end{equation}
Consequently one obtains the following set of differential equations 
\begin{equation}\label{ded}
\begin{pmatrix}
\dot a_1(t) \\
\dot a_2(t)
\end{pmatrix}=
-igf(t)
\underbrace{\begin{pmatrix}
0 & 1 \\
1 & 0
\end{pmatrix}}_{\tau_{x}}
\begin{pmatrix}
a_1(t) \\
a_2(t)
\end{pmatrix}
\end{equation}
where $\tau_{x}$ denotes the $x$ direction Pauli matrices in the two-dimensional vector space spanned by $|\psi_{1}\rangle$ and $|\psi_{2}\rangle$. This matrix can be diagonalized by the unitary transformation \begin{equation}
\label{Ud} U=
\begin{pmatrix}
-\dfrac{1}{\sqrt{2}} & \dfrac{1}{\sqrt{2}}\\\\
\dfrac{1}{\sqrt{2}} & \dfrac{1}{\sqrt{2}}
\end{pmatrix} 
\end{equation}
with eigenstates $|\psi_{\pm}\rangle$.
It follows that Equation~(\ref{ded}) can be rewritten as a set of decoupled differential equations
\begin{eqnarray}
\dot a_-(t)&=&igf(t) a_-(t) \\
\dot a_+(t)&=&-igf(t) a_+(t).
\end{eqnarray}
with 
\begin{equation}
a_{\pm}=\frac{a_{2}\pm a_{1}}{\sqrt{2}}
\end{equation}
and solutions
\begin{eqnarray}
a_-(t)&=&a_-(0)e^{ig\int_0^t dt'f(t')}  \\
a_+(t)&=&a_+(0)e^{-ig\int_0^t dt' f(t')}.
\end{eqnarray}
Introducing the time dependent angle
\begin{equation}
\label{varphi}
\varphi(t)=g\int_0^t dt'f(t').
\end{equation}
It follows that in the new basis the time evolved state $|\psi(t)\rangle$ can be written as
\begin{equation}
|\psi(t)\rangle=a_{-}(t) |\psi_{-}\rangle+a_{+}(t)|\psi_{+}\rangle\equiv
\begin{pmatrix}
a_-(0)e^{i\varphi(t)} \\
a_+(0)e^{-i\varphi(t)}
\end{pmatrix}.
\end{equation}

%\textcolor{red}{[toglierei questi passaggi] In the current basis, using the unitary transformation $U$ in Equation~(\ref{Ud}), the initial conditions read
%\begin{equation}
%\begin{pmatrix}
%a_-(0) \\
%a_+(0)
%\end{pmatrix}=
%\begin{pmatrix}
%\dfrac{1}{\sqrt{2}} & -\dfrac{1}{\sqrt{2}}\\\\
%\dfrac{1}{\sqrt{2}} & \dfrac{1}{\sqrt{2}}
%\end{pmatrix}
%\begin{pmatrix}
%1 \\
%0
%\end{pmatrix}=
%\begin{pmatrix}
%-\dfrac{1}{\sqrt{2}} \\\\
%\dfrac{1}{\sqrt{2}}
%\end{pmatrix}.
%\end{equation}
%Consequently the time evolved state becomes
%\begin{equation}|\psi(t)\rangle=
%\dfrac{1}{\sqrt{2}}\begin{pmatrix}
%-e^{i\varphi(t)} \\
%e^{-i\varphi(t)}
%\end{pmatrix}.
%\end{equation}}

Returning back to the initial basis by using the inverse transformation $U^{-1}$ one finally~obtains
\begin{equation}
\label{tsd}
|\psi(t)\rangle=\cos\varphi(t) |\psi_{1}\rangle-i \sin\varphi(t)|\psi_{2}\rangle\equiv
\begin{pmatrix}
\cos \varphi(t) \\\\
-i\sin \varphi(t)
\end{pmatrix}.
\end{equation}

Notice that, limited to the particular choice of the initial state in Equation~(\ref{insd}) the results derived previously holds also outside the RWA. 

%%%%%%%%%%%%%%%%%%%%%%%%%%%%%%

\subsection{Coupling Mediated by a Third TLS}
\label{TLS_mediated}

\begin{figure}[!]
\centering
\includegraphics[scale=0.4]{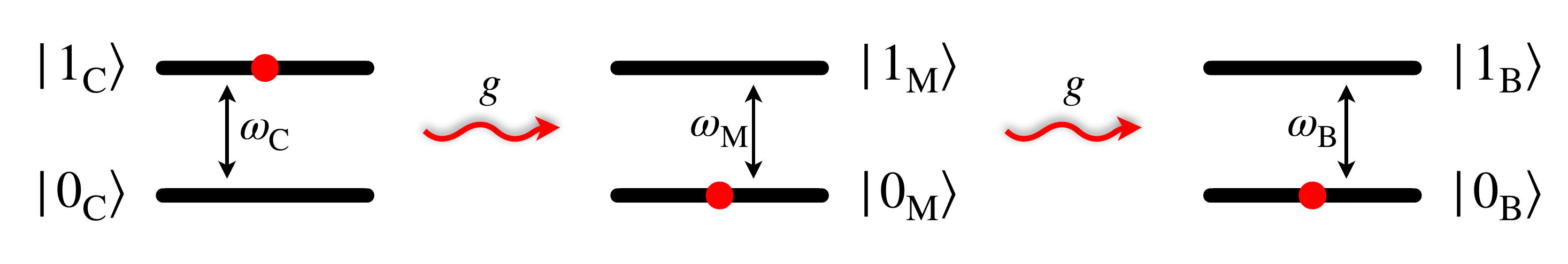} \vspace{-6pt}
 \caption {Scheme of a TLS-mediated energy transfer process, where the charger, with level spacing $\omega_{\rm C}$, the mediator, with level spacing $\omega_{\rm M}$, and the QB, with level spacing $\omega_{\rm B}$, are coupled via the same coupling constant $g$ which is further modulated through a time dependent switching function.}
   \label{fig2}
\end{figure}

Let's consider now a situation in which C and B are not directly coupled, but where the energy transfer is allowed by an intermediate TLS coupled with both of them (labelled with the index M), as in Figure~\ref{fig2}.  The Hamiltonian of the system in the RWA approximation is then given by  
\begin{equation}\label{Hmed} H_{\rm{med}}(t)= H_{0}+\frac{\omega_{\rm M}}{2}\sigma_z^{\rm M}+gf_{\rm CM}(t)(\sigma_-^{\rm C}\sigma_+^{\rm M}+\sigma_+^{\rm C}\sigma_-^{\rm M})+gf_{\rm BM}(t)(\sigma_-^{\rm B}\sigma_+^{\rm M}+\sigma_+^{\rm B}\sigma_-^{\rm M}), \end{equation}
where $\omega_{\rm M}$ is the level spacing of the mediator and $f_{\rm CM}(t)$ and $f_{\rm BM}(t)$ are two different time-dependent functions that will be specified later.

Extending the notation used in the previous Subsection, we consider as the initial state ($t=0$) of the total system  
\begin{equation}\label{insm}
|\Psi(0)\rangle=|1_{\rm C}0_{\rm M}0_{\rm B}\rangle\equiv|\Psi_{1}\rangle=\begin{pmatrix} 1 \\ 0\\ 0 \end{pmatrix}. 
\end{equation} 

Taking into account again the overall conservation of the number of excitations it is possible to use the basis
\begin{eqnarray}
|1_{\rm C}0_{\rm M}0_{\rm B}\rangle&\equiv&|\Psi_{1}\rangle=\begin{pmatrix} 1 \\ 0\\ 0 \end{pmatrix} \\\nonumber |0_{\rm C}1_{\rm M}0_{\rm B}\rangle&\equiv&|\Psi_{2}\rangle=\begin{pmatrix} 0 \\ 1\\ 0 \end{pmatrix} \\\nonumber |0_{\rm C}0_{\rm M}1_{\rm B}\rangle&\equiv&|\Psi_{3}\rangle=\begin{pmatrix} 0 \\ 0\\ 1 \end{pmatrix}.
\end{eqnarray}
Therefore, with the chosen initial state in Equation~(\ref{insm}) the Hamiltonian can be rewritten as a $3 \times 3$ matrix, whose elements are
\begin{equation} H_{ij}(t)=\langle \Psi_i|H_{\rm{med}}(t)|\Psi_j\rangle, \end{equation}
where $i,j=1,2,3$. The Hamiltonian can be then written as
\begin{equation}
H_{\rm {med}}(t)=
\begin{pmatrix}
\dfrac{\omega_{\rm C}-\omega_{\rm M}-\omega_{\rm B}}{2} & gf_{\rm CM}(t) & 0 \\\\
gf_{\rm CM}(t) & \dfrac{-\omega_{\rm C}+\omega_{\rm M}-\omega_{\rm B}}{2} & gf_{\rm BM}(t) \\\\
0 & gf_{\rm BM}(t) & \dfrac{-\omega_{\rm C}-\omega_{\rm M}+\omega_{\rm B}}{2}
\end{pmatrix}.
\end{equation}
To further simplify the discussion it is possible to consider the three TLSs to be identical ($\omega_{\rm C}=\omega_{\rm M}=\omega_{\rm B}$), for which the Hamiltonian assumes the form
\begin{equation}
H_{\rm {med}}(t)=
\begin{pmatrix}
-\dfrac{\omega_{\rm B}}{2} & gf_{\rm CM}(t) & 0 \\\\
gf_{\rm CM}(t) & -\dfrac{\omega_{\rm B}}{2} & gf_{\rm BM}(t) \\\\
0 & gf_{\rm BM}(t) & -\dfrac{\omega_{\rm B}}{2}
\end{pmatrix}, 
\end{equation}
where the constant diagonal term plays no role in the following analysis and will be therefore neglected. 

The Schr\"odinger equation now reads
\begin{equation}
i \frac{d}{dt}|\Psi(t)\rangle=H_{\rm{med}}(t)|\Psi(t)\rangle,
\end{equation}
where the time evolved state can be written in the spinorial notation as
\begin{equation} \label{psimed}
|\Psi(t)\rangle=\sum^{3}_{k=1}b_{k}(t)|\Psi_{k}\rangle\equiv 
\begin{pmatrix}
b_{1}(t) \\
b_{2}(t) \\
b_{3}(t)
\end{pmatrix}.
\end{equation}
We thus get the following set of differential equations
\begin{equation}\label{sein}
\begin{pmatrix}
\dot b_1(t) \\
\dot b_2(t) \\
\dot b_3(t)
\end{pmatrix}=-i
\begin{pmatrix}
 0 & gf_{\rm CM}(t) & 0 \\
gf_{\rm CM}(t) & 0 & gf_{\rm BM}(t) \\
0 & gf_{\rm BM}(t) & 0
\end{pmatrix}
\begin{pmatrix}
b_1(t) \\
b_2(t) \\
b_3(t)
\end{pmatrix}.
\end{equation}
In general, a numerical analysis is required in order to fully describe the dynamics of this system. However, two relevant limiting cases can be analytically solved.

\begin{figure}[h!]
\centering
\includegraphics[scale=0.4]{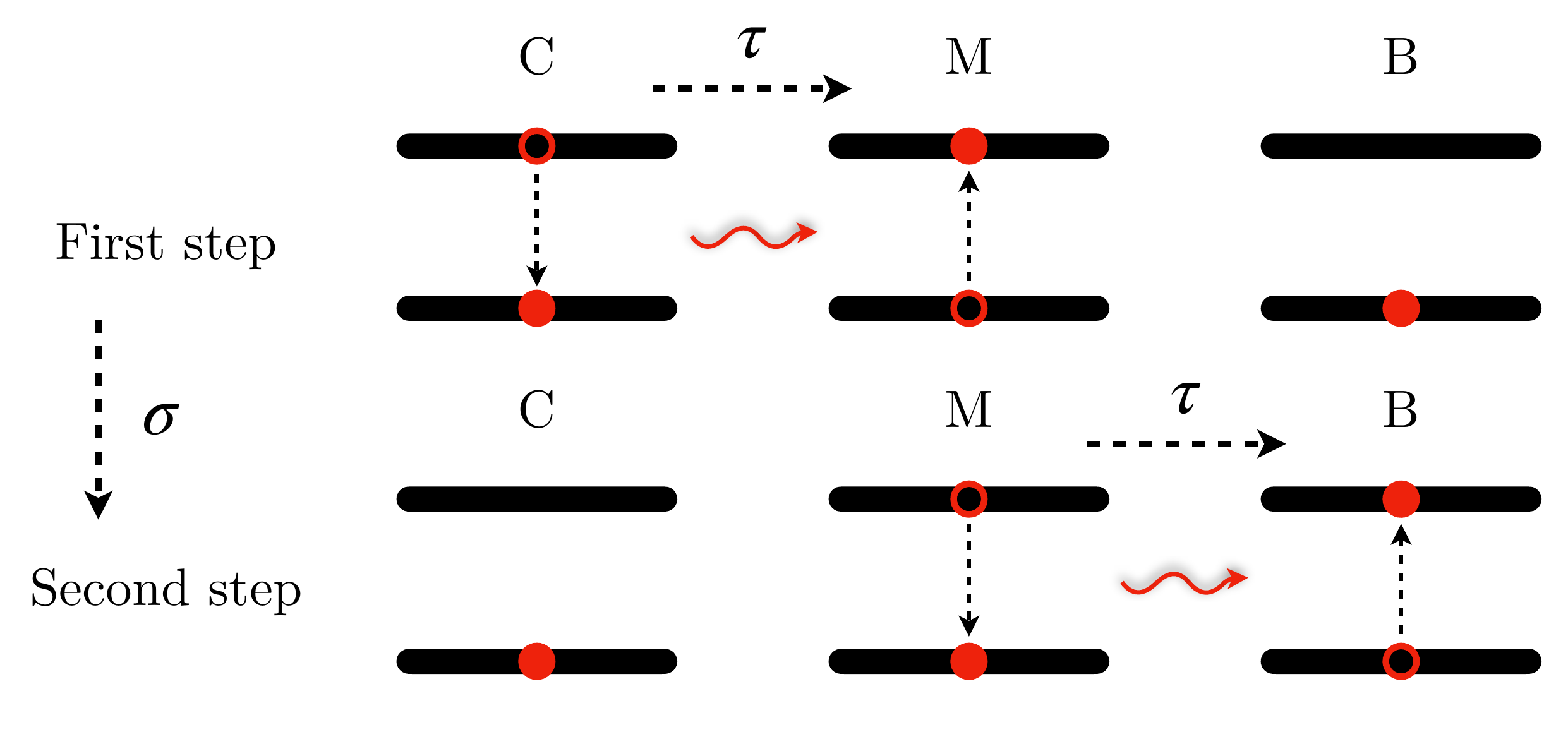} \vspace{-6pt}
 \caption {Scheme of the two-step TLS mediated energy transfer process. In the first step the energy stored in the charger is transferred to the mediator turning on the interaction between C and M for a finite time $\tau$. After a time $\sigma\gg \tau$ the mediator transfers this energy to the QB turning on the interaction between M and B again for a time $\tau$.}
   \label{fig2a}
\end{figure}

\begin{itemize}
    \item $f_{\rm BM}(t)=f_{\rm CM}(t-\sigma)$ (see Figure~\ref{fig2a}), with $\sigma \gg \tau$, $\tau$ being the typical time width in which the time-dependent functions are different from zero.

For $t\ll\sigma$ one gets
\begin{equation}\label{S1}
\begin{pmatrix}
\dot b_1(t) \\
\dot b_2(t) \\
\dot b_3(t)
\end{pmatrix}=-i
\begin{pmatrix}
 0 & gf_{\rm CM}(t) & 0 \\
gf_{\rm CM}(t) & 0 & 0 \\
0 & 0 & 0
\end{pmatrix}
\begin{pmatrix}
b_1(t) \\
b_2(t) \\
b_3(t)
\end{pmatrix},
\end{equation}
which describes an energy transfer between the charger and the mediator ($|1_{\rm C}0_{\rm M}0_{\rm B}\rangle \rightarrow |0_{\rm C}1_{\rm M}0_{\rm B}\rangle$). Instead for $t \sim \sigma$ the set of differential equations reads
\begin{equation}\label{S2}
\begin{pmatrix}
\dot b_1(t) \\
\dot b_2(t) \\
\dot b_3(t)
\end{pmatrix}=-i
\begin{pmatrix}
 0 & 0 & 0 \\
0 & 0 & gf_{\rm BM}(t) \\
0 & gf_{\rm BM}(t) & 0
\end{pmatrix}
\begin{pmatrix}
b_1(t) \\
b_2(t) \\
b_3(t)
\end{pmatrix},
\end{equation}
which represents an energy transfer between the mediator and the QB ($|0_{\rm C}1_{\rm M}0_{\rm B}\rangle \rightarrow |0_{\rm C}0_{\rm M}1_{\rm B}\rangle$).

This clearly describes a two-step energy transfer protocol where each step has the same form of the direct coupling case. In particular in the first step one has 
\begin{equation}
|\Psi(t)\rangle=\cos\varphi_{\mathrm{CM}}(t) |\Psi_{1}\rangle-i \sin\varphi_{\mathrm{CM}}(t)|\Psi_{2}\rangle=
\begin{pmatrix}
\cos \varphi_{\mathrm{CM}}(t) \\
-i\sin \varphi_{\mathrm{CM}}(t) \\
0
\end{pmatrix}
\end{equation}
with
\begin{equation}
\label{varphi}
\varphi_{\mathrm{CM}}(t)=g\int_0^t dt'f_{\mathrm{CM}}(t').
\end{equation}

Assuming a complete state (and energy) transfer form the charger to the mediator at the first step (see below for more details) the second step can be written as  

\begin{equation}
|\Psi(t)\rangle=\cos\varphi_{\mathrm{BM}}(t) |\Psi_{2}\rangle-i \sin\varphi_{\mathrm{BM}}(t)|\Psi_{3}\rangle=
\begin{pmatrix}
0 \\
\cos \varphi_{\mathrm{BM}}(t) \\
-i\sin \varphi_{\mathrm{BM}}(t)
\end{pmatrix}
\end{equation}
with 
\begin{equation}
\label{varphi}
\varphi_{\mathrm{BM}}(t)=g\int_0^t dt'f_{\mathrm{BM}}(t').
\end{equation}

In this process the energy remains trapped into the mediator for a time of the order of $\sigma$. Therefore, this protocol can be considered as realistic as long as $\sigma$ is shorter with respect to the typical dephasing and relaxation times of the mediator~\cite{Carrega20, Caravelli21}.

\item $f_{\rm CM}(t)=f_{\rm BM}(t)=f(t)$.
This type of protocol describes a simultaneous transfer from the charger to the mediator and to the QB. 

In this case the set of differential equations in Equation~(\ref{sein}) reduces to
\begin{equation}\label{sde}
\begin{pmatrix}
\dot b_1(t) \\
\dot b_2(t) \\
\dot b_3(t)
\end{pmatrix}=-igf(t)
\underbrace{\begin{pmatrix}
 0 & 1 & 0 \\
1 & 0 & 1 \\
0 & 1 & 0
\end{pmatrix}}_{\mathcal{T}}
\begin{pmatrix}
b_1(t) \\
b_2(t) \\
b_3(t)
\end{pmatrix}.
\end{equation}
The matrix $\mathcal{T}$ is diagonalized by the unitary matrix 
\begin{equation}\label{U}
\mathcal{U}=\begin{pmatrix}
\dfrac{1}{2} & -\dfrac{1}{\sqrt{2}} & \dfrac{1}{2} \\\\
\dfrac{1}{2} & \dfrac{1}{\sqrt{2}} & \dfrac{1}{2} \\\\
-\dfrac{1}{\sqrt{2}} & 0 & \dfrac{1}{\sqrt{2}}
\end{pmatrix}.
\end{equation}

This allows to rewrite Equation~(\ref{sde}) as a set of decoupled equations
\begin{eqnarray}
\dot b_-(t)&=& \sqrt{2}igf(t)b_-(t) \\\nonumber
\dot b_+(t)&=& -\sqrt{2}igf(t)b_+(t) \\\nonumber
\dot b_{0}(t)&=&0,
\end{eqnarray}
with 
\begin{eqnarray}
b_{-}&=&\frac{1}{2}b_{1}-\frac{1}{\sqrt{2}}b_{2}+\frac{1}{2}b_{3}\\\nonumber
b_{+}&=&\frac{1}{2}b_{1}+\frac{1}{\sqrt{2}}b_{2}+\frac{1}{2}b_{3}\\\nonumber
b_{0}&=&-\frac{1}{\sqrt{2}}b_{1}+\frac{1}{\sqrt{2}}b_{3}
\end{eqnarray}

and general time evolution
\begin{eqnarray}
b_-(t)&=&b_-(0) e^{\sqrt{2}ig\int_0^t dt' f(t')} \\\nonumber
b_+(t)&=&b_+(0) e^{-\sqrt{2}ig\int_0^t dt' f(t')} \\\nonumber
b_{0}(t)&=&b_{0}(0).
\end{eqnarray}
In this basis the time evolved state $|\Psi(t)\rangle$ is
\begin{equation}
|\Psi(t)\rangle=\begin{pmatrix}
b_-(0)e^{i\phi(t)} \\
b_+(0)e^{-i\phi(t)} \\
b_{0}(0)
\end{pmatrix},
\end{equation}
with
\begin{equation}\label{phi}
\phi(t)=\sqrt{2}g\int_0^t dt' f(t').
\end{equation}

%\textcolor{red}{[toglierei questi passaggi] The initial conditions in the current basis can be found using the unitary transformation $\mathcal{U}$ in such a way that
%\begin{equation}
%\begin{pmatrix}
%b_-(0) \\
%b_+(0) \\
%b_{0}(0)
%\end{pmatrix}= 
%\begin{pmatrix}
%\dfrac{1}{2} & -\dfrac{1}{\sqrt{2}} & \dfrac{1}{2} \\\\
%\dfrac{1}{2} & \dfrac{1}{\sqrt{2}} & \dfrac{1}{2} \\\\
%-\dfrac{1}{\sqrt{2}} & 0 & \dfrac{1}{\sqrt{2}}
%\end{pmatrix}
%\begin{pmatrix}
%1 \\
%0 \\
%0
%\end{pmatrix}=
%\begin{pmatrix}
%\dfrac{1}{2} \\\\
%\dfrac{1}{2} \\\\
%-\dfrac{1}{\sqrt{2}}
%\end{pmatrix}.
%\end{equation}}
%\textcolor{red}{Consequently the time evolved state becomes
%\begin{equation}
%|\Psi(t)\rangle=\begin{pmatrix}
%\dfrac{1}{2}e^{i\phi(t)} \\\\
%\dfrac{1}{2}e^{-i\phi(t)} \\\\
%-\dfrac{1}{\sqrt{2}}
%\end{pmatrix}.
%\end{equation}}

By using the inverse transformation $\mathcal{U}^{-1}$ it is possible to rewrite it in the original basis as follows 
\centering
\begin{equation}\label{psi}
|\Psi(t)\rangle=\dfrac{1}{2}[\cos \phi(t) +1]|\Psi_{1}\rangle-\dfrac{i}{\sqrt{2}}\sin \phi(t)|\Psi_{2}\rangle+\dfrac{1}{2}[\cos \phi(t) - 1]|\Psi_{3}\rangle\equiv
\begin{pmatrix}
\dfrac{1}{2}[\cos \phi(t) +1] \\\\
-\dfrac{i}{\sqrt{2}}\sin \phi(t) \\\\
\dfrac{1}{2}[\cos \phi(t) - 1]
\end{pmatrix}.
\end{equation}
\end{itemize}  

Also in this case, for the initial condition in Equation~(\ref{insm}) the results derived in this Subsection can be extended outside the RWA.
\vspace{-6pt}
 
\section{Stored Energy in the TLSs}\label{Energy}
To properly characterize the energy transfer between C and B it is necessary to consider the time evolution of the stored energy in each TLS composing the device in the different configurations introduced above.

\subsection{Direct Coupling Case}

The stored energy inside the QB at a given time is defined as
\begin{equation}\label{Eb} E_{\rm B}(t)\equiv \langle \psi(t) | H_{\rm B} | \psi(t) \rangle- \langle \psi(0) | H_{\rm B} | \psi(0) \rangle. \end{equation}

Taking into account the time evolved state in Equation~(\ref{tsd}) and the initial condition $| \psi(0) \rangle=|1_{\rm C}0_{\rm B}\rangle$, one gets
\begin{equation} 
\label{Ebd}
E_{\rm B}(t)=
\omega_{\rm B}|a_{2}(t)|^2 = \omega_{\rm B}\sin^2 \varphi(t).
\end{equation}
It is also possible to introduce the first maximum of the stored energy
\begin{equation}\label{Ebmax} E_{\rm B,max}\equiv E_{\rm B}(t_{\rm B,max}), \end{equation}
where $t_{\rm B,max}$ represents the shorter transfer time.

Concerning the charger one can write
\begin{equation}\label{Ec} 
E_{\rm C}(t)\equiv \langle \psi(t) | H_{\rm C} | \psi(t) \rangle- \langle \psi(0) | H_{\rm C} | \psi(0) \rangle, \end{equation}
namely
\begin{equation}\label{Ecd}
E_{\rm C}(t)=-\omega_{\rm B} |a_{2}(t)|^2= -\omega_{\rm B}\sin^2 \varphi(t),
\end{equation}
which is negative consistently with the fact that C releases energy towards B. In particular, it is possible to define
\begin{equation} E_{\rm C,max}\equiv E_{\rm C}(t_{\rm B,max}), \end{equation}
the value assumed by the stored energy in the charger at the transfer time $t_{\rm B,max}$. Notice that, since the system is considered on-resonance, the maxima of the stored energy in the QB coincide with the minimum of the energy stored in the charger. Moreover, from Equations~(\ref{Ebd}) and~(\ref{Ecd}), one obtains that for all times 
\begin{equation}
E_{\rm B}(t)+E_{\rm C}(t)=0, \end{equation} 
proving the energy conservation in the charger-QB system. As demonstrated in \cite{Andolina18, Crescente22}, when the system is considered on-resonance ($\omega_{\rm C}=\omega_{\rm B}$), no work is necessary to switch on and off the interaction, meaning that the interaction energy is also null $E_{\rm int}(t)=0$.  

\subsection{TLS-Mediated Case}

Here, for both the considered cases in Section~\ref{TLS_mediated} the stored energy in the QB and charger, starting from the initial condition $|1_{\rm C}0_{\rm M}0_{\rm B}\rangle$, are described by the general relations
\begin{eqnarray}
\label{Ett} 
E_{\rm B}(t)&=& \omega_{\rm B}|b_3(t)|^2 \nonumber\\ E_{\rm C}(t)&=& -\omega_{\rm B}\bigg[|b_2(t)|^2+|b_3(t)|^2\bigg]. 
\end{eqnarray}

In presence of a mediator it is also useful to evaluate the energy stored inside it as 
\begin{equation}\label{Em}
E_{\rm M}(t)\equiv \langle \Psi(t) | H_{\rm M} | \Psi(t) \rangle- \langle \Psi(0) | H_{\rm M} | \Psi(0) \rangle, \end{equation}
with 
\begin{equation}
H_{\rm M}=\frac{\omega_{\rm M}}{2}\sigma_z^{\rm M}.
\end{equation}
Consequently, one gets the general relation
\begin{equation}\label{Emt}  E_{\rm M}(t)=\omega_{\rm B}|b_2(t)|^2.\end{equation}

Notice that, from the Equations~(\ref{Ett}) and~(\ref{Emt}) the energy of the system is conserved for all times, namely 
\begin{equation}
E_{\rm B}(t)+E_{\rm C}(t)+E_{\rm M}(t)=0.
\end{equation} 
Also in this case, this is a consequence of the considered initial condition and of the resonance among the TLSs.

\begin{itemize}

\item For $f_{\rm BM}(t)=f_{\rm CM}(t-\sigma)$ one obtains
\begin{eqnarray}\label{Ebmstep}E_{\rm B}(t)&=&\omega_{\rm B}\sin^2 \varphi_{\rm BM}(t) \\
E_{\rm C}(t)&=&-\omega_{\rm B}\sin^2 \varphi_{\rm CM}(t) \\
\label{Emmstep}E_{\rm M}(t)&=&\omega_{\rm B}[\sin^2\varphi_{\rm CM}(t)-\sin^2 \varphi_{\rm BM}(t)],
\end{eqnarray}
where
\begin{equation}\label{anglem}
\varphi_{\rm BM}(t)=g\int_0^t dt'f_{\rm BM}(t') \quad\quad \varphi_{\rm CM}(t)=g\int_0^t dt'f_{\rm CM}(t').\end{equation}

\item For $f_{\rm CM}(t)=f_{\rm BM}(t)=f(t)$ the stored energy inside the QB as function of $\phi(t)$ is given by
\begin{equation}\label{Ebm}
E_{\rm B}(t)=\omega_{\rm B}\bigg[\frac{1}{2}\cos \phi(t) -\frac{1}{2}\bigg]^2. \end{equation}
In the same way, from Equation~(\ref{Ett}) it is possible to obtain the energy of the charger which reads
\begin{equation}
\label{Ecm}
E_{\rm C}(t)=\omega_{\rm B}\bigg[\frac{1}{4}\cos^2\phi(t) +\frac  {1}{2}\cos \phi(t)- \frac{3}{4}\bigg]. 
\end{equation}
While for the mediator one obtains
\begin{equation}\label{Emm}
E_{\rm M}(t)=\frac{\omega_{\rm B}}{2}\sin^2\phi(t). \end{equation}

These expressions allow to evaluate the maximum of the energy stored inside the QB, as in Equation~(\ref{Ebmax}), the corresponding charging time and also the value assumed by $E_{\rm C}(t)$ and  $E_{\rm M}(t)$ when the QB reaches its maximum. 

\end{itemize}

\section{Switching Function}\label{FF}

The forms of the functions in Equations~(\ref{H_dir}) and~(\ref{Hmed}) are now specified. As discussed above, ideally in both cases these functions need to be (almost) zero everywhere except for a window of width $\sim\tau$ in time where they saturates (close) to one. This is done in order to switch on and off of the interaction between the TLSs. Moreover, the value of $\tau$ is chosen in such a way to optimize the energy transfer~\cite{Crescente22}. 
It is worth to mention that such kind of situation can be implemented experimentally in superconducting circuits acting on the capacitive coupling between transmons playing the role of TLSs~\cite{Krantz19}. In the following a step-wise function is chosen as the simplest possible way to fulfill the above requirements even if smoother profiles can be considered~\cite{Crescente22, Thomas22}.

\subsection{Direct Coupling Case}

\begin{figure}[h!]
\centering
\includegraphics[scale=0.45]{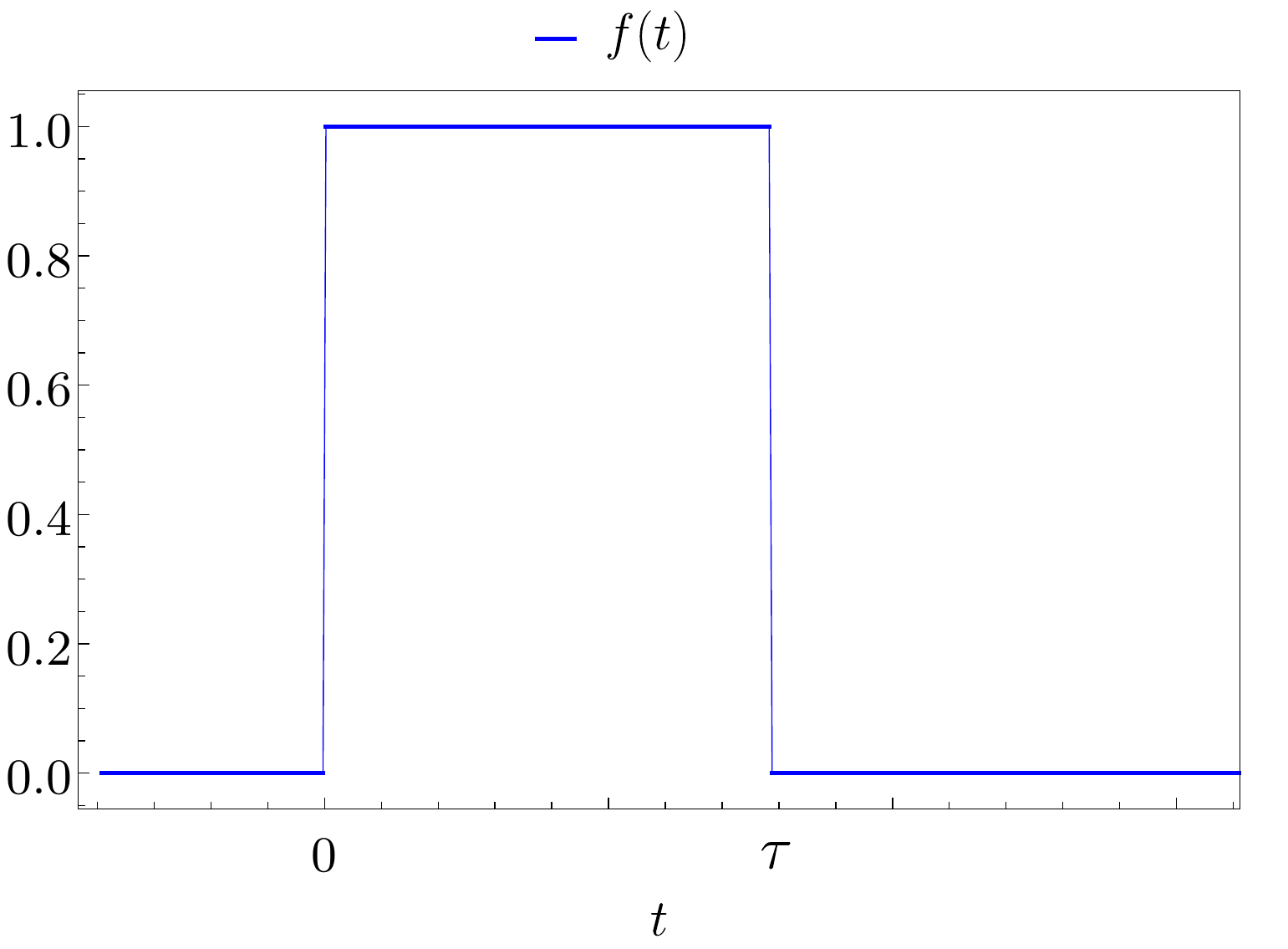} 
 \caption {Behaviour of $f(t)$ as function of time $t$, switched on for a time interval $\tau$.}
   \label{fig3}
\end{figure}

For a direct transfer scenario, we choose a switching on and off function of the form (see Figure~\ref{fig3})
\begin{equation} \label{ft}
f(t)= \theta(t)-\theta(t-\tau)=
\begin{cases}
0 \quad \text{if} \ t<0 \\
1 \quad \text{if} \ 0\leq t<\tau \\
0 \quad \text{if} \ t\geq \tau
\end{cases},
\end{equation}
where $\theta(t)$ indicates the Heaviside step function and, as stated above, $\tau$ is the time interval for which the coupling is different from zero. Moreover, the time $\tau=t_{\rm B, max}$ is chosen in order to switch off the interaction exactly when the QB reaches the first maximum of the stored energy~\cite{Crescente22}.

In this case the angle $\varphi(t)$ in Equation~(\ref{varphi}) is
\begin{equation}\varphi(t)=
\begin{cases}
0 \quad \ \ \ \text{if} \ t<0 \\
gt \quad \ \text{if} \ 0\leq t<\tau \\
g\tau \quad  \text{if} \ t\geq \tau
\end{cases}.
\end{equation}
\vspace{-12pt}

\subsection{TLS-Mediated Coupling Case}

As discussed above, for a TLS-mediated transfer it is possible to identify two protocols. 

\begin{figure}[h!]
\centering
\includegraphics[scale=0.45]{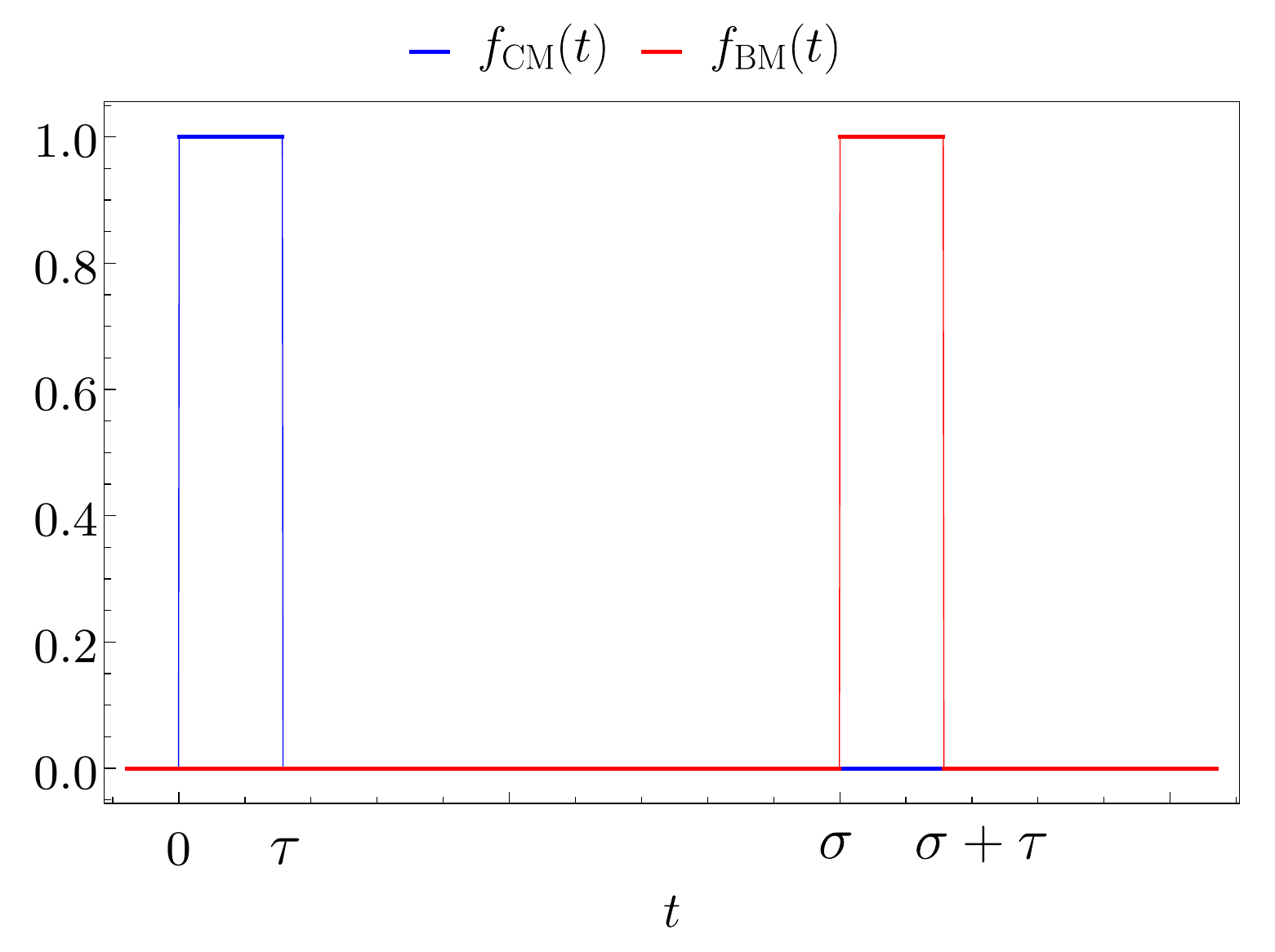} \vspace{-6pt}
 \caption {Behaviour of $f_{\rm CM}(t)$ (blue curve) and $f_{\rm BM}(t)$ (red curve) as function of time $t$. The two functions are switched on for a time $\tau$ and $\sigma$ represents the delay between them.}
   \label{fig4}
\end{figure}

In the two-step energy transfer, where $f_{\rm BM}(t)=f_{\rm CM}(t-\sigma)$, one possible choice to switch on and off the interaction is to assume $f_{\rm CM}(t)$ as in Equation~(\ref{ft}) and $f_{\rm BM}(t)$ shifted consequently (see Figure~\ref{fig4}).

In this case the values of the angles in Equation~(\ref{anglem}) are
\begin{eqnarray}
\label{varphistep} \varphi_{\rm CM}(t)&=&g\int_0^t dt'f_{\rm CM}(t')=
\begin{cases}
0 \quad \ \ \text{if} \ t<0 \\
gt \quad \ \ \text{if} \ 0\leq t<\tau \\
g \tau \quad \text{if} \ t\geq \tau
\end{cases} \nonumber \\
\varphi_{\rm BM}(t)&=&g\int_0^t dt'f_{\rm BM}(t')=
\begin{cases}
0 \quad \ \ \text{if} \ t<\sigma \\
gt \quad \text{if} \ \sigma\leq t<\tau+\sigma \\
g\tau \quad \text{if} \ t\geq \tau+\sigma
\end{cases}.
\end{eqnarray}

Different is the situation when $f_{\rm CM}(t)=f_{\rm BM}(t)=f(t)$ [see Equation~(\ref{ft})]. Here, from Equation~(\ref{phi}) the angle $\phi(t)$ becomes
\begin{equation}\phi(t)=\sqrt{2}g\int_0^t dt' f(t')=
\begin{cases}
0 \qquad\quad \text{if} \ t<0 \\
\sqrt{2}gt \quad \ \text{if} \ 0\leq t<\tau \\
\sqrt{2}g\tau \quad  \text{if} \ \ t\geq \tau
\end{cases}.
\end{equation}
\vspace{-13pt}

\subsection{Analytical form of the Energy Stored in the QB and of the Relative Transfer Time}

To understand how fast it is possible to transfer energy from the charger to the QB it is necessary to rewrite the stored energy inside the QB and to determine the corresponding time required to transfer this energy. This can be done analytically starting from the switching functions just introduced. 

For the direct scenario, starting from Equation~(\ref{Ebd}), the stored energy inside the QB (for $t$ such that $0\leq t<\tau$) becomes
\begin{equation}E_{\rm B}(t)= \omega_{\rm B}\sin^2 \varphi(t)=\omega_{\rm B}\sin^2 gt. \end{equation}
As a consequence the maximum of the stored energy $E_{\rm B, max}=\omega_{\rm B}$ is obtained for times such that 
\begin{equation}
\label{tmaxd}t_{\rm B, max}=\frac{k\pi}{2g}, \end{equation}
where $k \in \mathbb{Z}$.

A similar consideration can be done for the two steps mediated charging protocol, where the charging time is the same evaluated above plus a controlled delay given by $\sigma$. According to this, one has
\begin{equation}
\label{tmaxs} t_{\rm B, max}=\frac{k\pi}{2g}+\sigma. \end{equation}

Instead, for the TLS-mediated case, from Equation~(\ref{Ebm}) one gets
\begin{equation}
E_{\rm B}(t)=\frac{\omega_{\rm B}}{2}\bigg[\frac{1}{\sqrt{2}}\cos \sqrt{2}gt -\frac{1}{\sqrt{2}}\bigg]^2. \end{equation}
Here, the maximum of the stored energy $E_{\rm B, max}=\omega_{\rm B}$ is obtained for times
\begin{equation}\label{tmaxm} \sqrt{2}gt= (2k+1)\pi \quad \Rightarrow \quad t_{\rm B, max}=\frac{(2k+1)\pi}{\sqrt{2}g}, \end{equation}
with $k \in \mathbb{Z}$.

%%%%%%%%%%%%%%%%%%%%%%%%%

\section{Results}\label{results}

In this Section, the analytical results discussed above for the direct, two-step and coherent TLS-mediated scenarios are reported to determine their energy transfer performances. The validity of the analytical calculations is further checked through exact diagonalization (see Ref.~\cite{Crescente22} for more details).
    
\subsection*{Direct vs. TLS-Mediated Scenarios}

As a starting point we consider the direct coupling case and the results concerning the energies stored in the different parts of the system, are reported for the representative coupling constant $g=0.05~\omega_{\rm B}$. 

\begin{figure}[h!]
\centering
\includegraphics[scale=0.5]{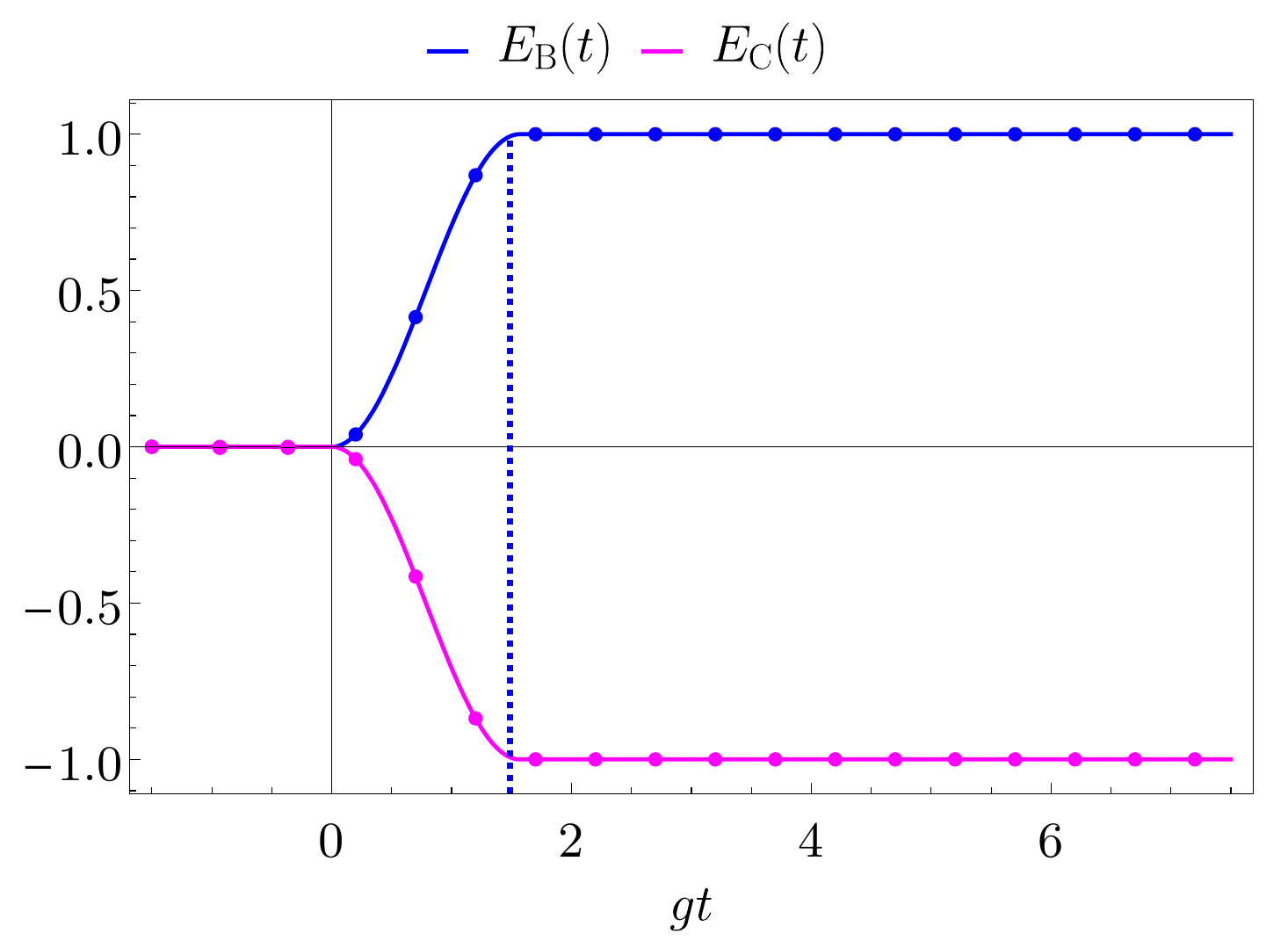} \vspace{-6pt}
 \caption {Behaviour of $E_{\rm B}(t)$ (blue curve) and $E_{\rm C}(t)$ (magenta curve) in units of $\omega_{\rm B}$ as a function of $g t$ with $g=0.05 \omega_{\rm B}$ for the direct coupling case. The value $g\tau=\pi/2$ is considered to switch off the interaction when the first maximum of the transferred energy is achieved (dashed blue vertical line). Dots on the curves represent the numerical results obtained by exact diagonalization.}
   \label{fig5}
\end{figure}

In Figure~\ref{fig5} it is possible to observe that the charger completely discharge decreasing its energy from $0$ to $-\omega_{\rm B}$, while the QB has the opposite behaviour, starting from being empty ($E_{\rm B}=0$) to the completely charged situation $E_{\rm B, max}=\omega_{\rm B}$. This proves the complete energy transfer in the direct scenario, obtained for
\begin{equation}
g\tau=gt_{\rm B, max}=\frac{\pi}{2},
\end{equation} 
as a consequence of Equation~(\ref{tmaxd}). Moreover, the numerical results (dots in Figure~\ref{fig5}) are in full agreement with the presented analytical model.

\newpage

\begin{figure}[h!]
\centering
\includegraphics[scale=0.5]{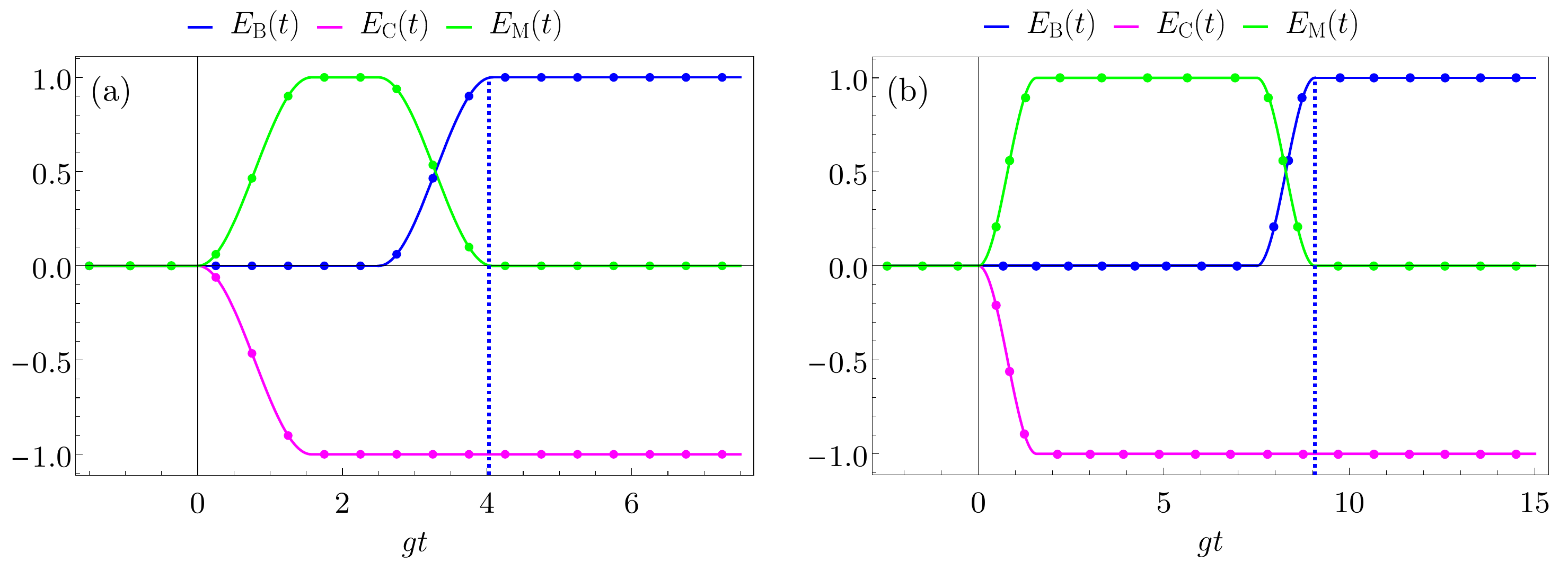} \vspace{-6pt}
 \caption {Behaviour of $E_{\rm B}(t)$ (blue curves), $E_{\rm C}(t)$ (magenta curves), $E_{\rm M}(t)$ (green curves) in units of $\omega_{\rm B}$ and as a function of $g t$ with $g=0.05 \omega_{\rm B}$ for the two step TLS-mediated case with $g\sigma=2.5$ (\textbf{a}) and $g\sigma=7.5$ (\textbf{b}). The values $g\tau=\pi/2+g\sigma$ are considered to switch off the interaction when the first maximum of the transferred energy is achieved (dashed blue vertical lines). Dots on the curves represent the numerical results obtained by exact diagonalization.}
   \label{fig6}
\end{figure}

We now focus on the TLS-mediated cases. Firstly, the two-step TLS-mediated case is reported in Figure~\ref{fig6} for two representative values of delay $g\sigma= 2.5$ [panel (a)] and $g\sigma= 7.5$ [panel (b)]. Here, it is possible to observe that initially, for times $t< \sigma$ the charger transfers its energy to the mediator and the energy stored into the QB remains null, as confirmed by Equations~(\ref{Ebmstep})--(\ref{Emmstep}) and~(\ref{varphistep}), where
\begin{eqnarray} \label{Estep1}
E_{\rm C}(t)&=&
\begin{cases}
-\omega_{\rm B}\sin^2 gt \ \quad 0\leq t<\tau \\
-\omega_{\rm B}\sin^2 g\tau \quad \tau \leq t < \sigma
\end{cases} \nonumber \\
E_{\rm M}(t)&=&
\begin{cases}
\omega_{\rm B}\sin^2 gt \ \ \ \ \quad 0\leq t<\tau \\
\omega_{\rm B}\sin^2 g\tau \ \ \ \quad \tau \leq t < \sigma
\end{cases} \nonumber \\ \nonumber \\
E_{\rm B}(t)&=&0 \ \ \ \quad\quad\quad\quad\quad\quad t \leq \sigma. \end{eqnarray}

Then, for a time $\sigma$ the energy remains stored in the mediator and available for the following transfer. 
Changing the value of $\sigma$ only modifies the time for which the mediator stays fully charged. As anticipated before, this time needs to be shorter compared to the typical decoherence and dephasing times of the system, in order to have the possibility to transfer all the energy in the QB.
For time $t>\sigma$ the mediator releases all its energy to the QB, allowing a complete transfer, while the energy in the charger remains constant. 
This is proved writing, again from Equations~(\ref{Ebmstep})--(\ref{Emmstep}) and~(\ref{varphistep}), the energy stored in the different parts of the system
\begin{eqnarray} \label{Estep2}
E_{\rm M}(t)&=&
\begin{cases}
\omega_{\rm B}(\sin^2g\tau-\sin^2 gt) \ \quad \sigma \leq t<\sigma+\tau \\
0 \qquad\qquad\qquad\qquad\qquad t\geq \sigma+\tau
\end{cases} \nonumber \\
E_{\rm B}(t)&=&
\begin{cases}
\omega_{\rm B}\sin^2 gt \ \qquad\qquad\qquad \sigma \leq t<\sigma+\tau \\
\omega_{\rm B}\sin^2 g\tau \qquad\qquad\qquad t\geq \sigma+\tau
\end{cases} \nonumber \\
 E_{\rm C}(t)&=&-\omega_{\rm B}\sin^2 g\tau \qquad\qquad\qquad t\geq \sigma. 
 \end{eqnarray} 
 
In this scenario, as a consequence of the delay time $\sigma$, the transfer time is longer compared to the direct case, and in particular, from Equation~(\ref{tmaxs}), it is $gt_{\rm B, max}\sim4.1$ (for $g\sigma=2.5$) and $gt_{\rm B, max}\sim9.1$ (for $g\sigma=7.5$).

\begin{figure}[h!]
\centering
\includegraphics[scale=0.5]{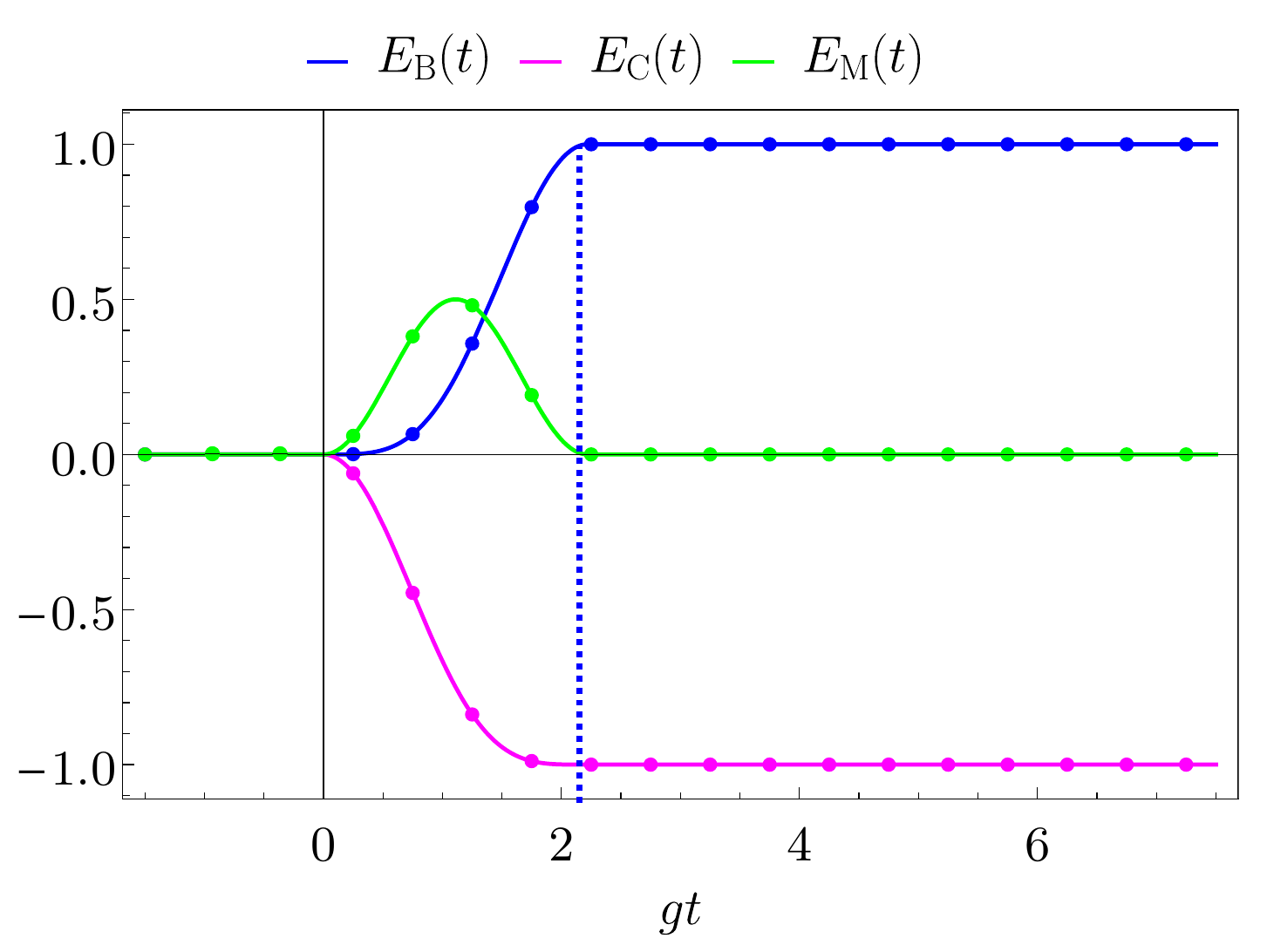} \vspace{-4pt}
 \caption {Behaviour of $E_{\rm B}(t)$ (blue curves), $E_{\rm C}(t)$ (magenta curves), $E_{\rm M}(t)$ (green curves) in units of $\omega_{\rm B}$ and as a function of $g t$ with $g=0.05 \omega_{\rm B}$ for the coherent TLS-mediated case. The value $g\tau=\pi/\sqrt{2}$ is considered to switch off the interaction when the first maximum of the transferred energy is achieved (dashed blue vertical line). Dots on the curves represent the numerical results.}
   \label{fig7}
\end{figure}

Finally, the simultaneous energy transfer is reported in Figure~\ref{fig7}. Here, it is possible to observe that the charger initially transfers its energy to the mediator. Only at a slightly later time the QB can extract the energy from the mediator, leading to a complete energy transfer. It is worth to underline the fact that in this configuration the mediator is at most half charged. Moreover, while this process can be faster compared to the two-step TLS-mediated scenario, it is in any case slower with respect to the direct one. In fact, starting from Equation~(\ref{tmaxm}), the transfer time is 
\begin{equation}
gt_{\rm B, max}=\frac{\pi}{\sqrt{2}}.   
\end{equation}

 Notice that also in this case the numerical and analytical results are perfectly in accord.

 %%%%%%%%%%%%%%%%%%%%%%%%%%%%%%

\begin{figure}[h!]
\centering
\includegraphics[scale=0.55]{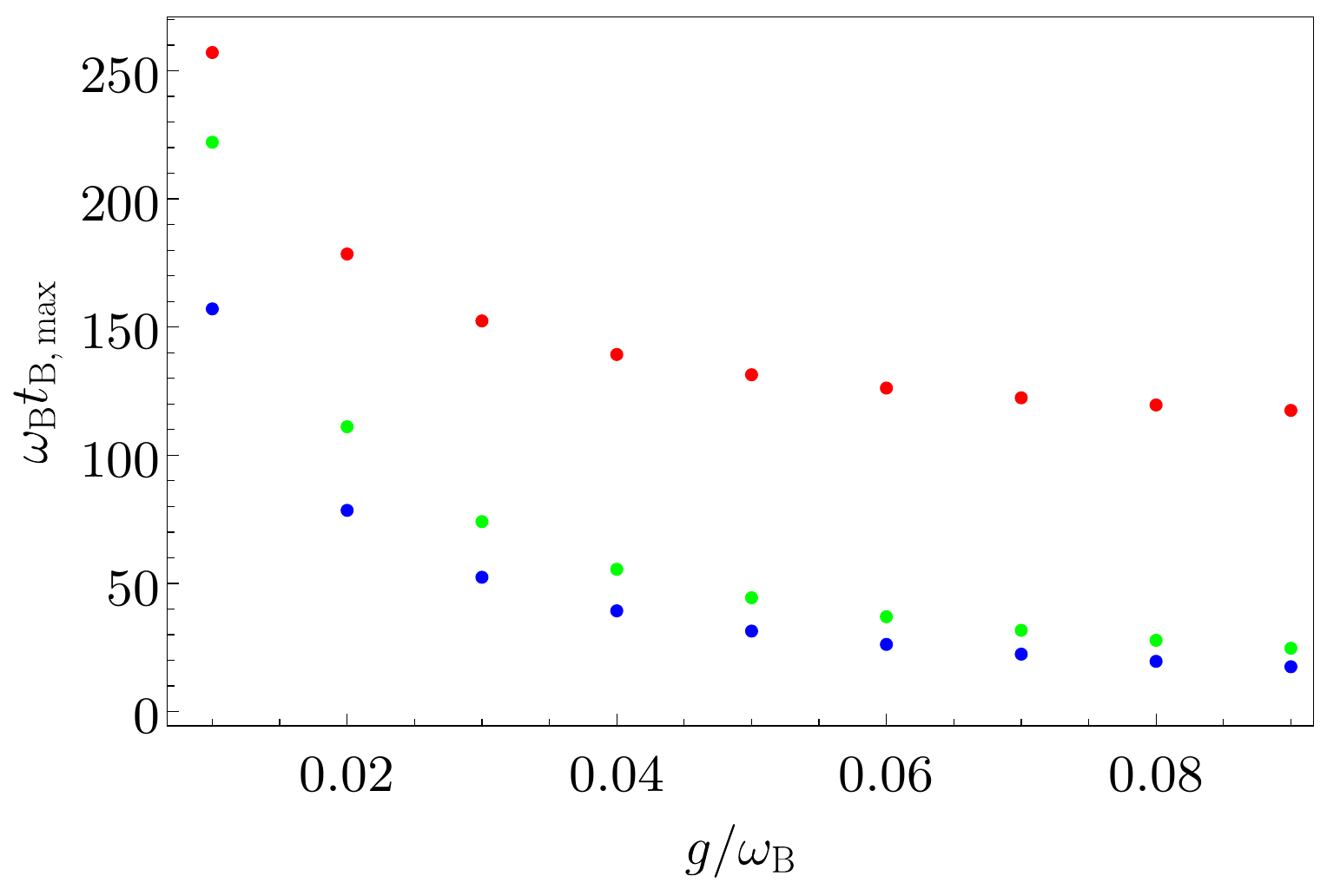} 
 \caption {Behaviour of $\omega_{\rm B}t_{\rm B,max}$ as function of $g/\omega_{\rm B}$ for the direct case (blue dots), the two-step TLS-mediated case with $\omega_{\rm B}\sigma=100$ (red dots) and for the coherent TLS-mediated case (green dots).}
   \label{fig8}
\end{figure}

In Figure~\ref{fig8} the behaviour of $\omega_{\rm B}t_{\rm B, max}$ is reported as a function of $g/\omega_{\rm B}$, in a range where the RWA is fulfilled.   
This plot further corroborate the behaviour of the transfer time in the three scenarios [see Equations~(\ref{tmaxd}), (\ref{tmaxs}) and~(\ref{tmaxm})], indicating the direct case as the fastest. However, increasing the value of $g$, the transfer time in the coherent TLS-mediated scenario becomes comparable to the direct one, but it never becomes faster. The two-step TLS-mediated case is obviously shifted of a time $\sigma$ with respect to the direct case.

To summarize, all the scenarios under investigation allow a complete energy transfer between the charger and the QB. Moreover, it is possible to state that the direct energy transfer is the fastest of the considered cases. However, the TLS-mediated cases open to the possibility to create more complex devices for energy transfer by progressively adding identical TLSs as building blocks. Moreover, the two-step TLS-mediated case allows to transfer the energy between a charger and a mediator with the possibility to store the energy for a certain time before release it on-demand to the QB.

\section{Conclusions}\label{conclusions}
The energy transfer process between a quantum charger and a quantum battery, modeled as identical two-level systems, has been considered. The direct transfer scenario is taken as a reference for the comparison with a situation where the transfer is mediated by an additional two-level system. In this latter configuration, it is possible to consider both a two-step process where the energy is firstly transferred from the charger to the mediator and then from the mediator to the battery and a case in which the two transfers occurs at the same time. The differences between these approaches are discussed by means of an analytically solvable model. The results have been also validated through exact diagonalization~\cite{Crescente22}. The main result is that the direct energy transfer process allows a faster transfer compared to the mediated cases. However, by increasing the value of the coupling constant, the coherent TLS-mediated scenario becomes comparable but never faster. Instead, the two-step mediated transfer allows a controllable energy transfer process, dictated by the chosen delay time in the switch on and off function of the interaction terms. 

This study represents a first step towards the investigation of a network devoted to the coherent transfer of energy for application to complex quantum devices. Moreover, the results considered here can be tested on nowadays quantum devices, where the coupling between the qubits can be controlled in time~\cite{Alexander20}.

%%%%%%%%%%%%%%%%%%%%%%%%%%%%%%%%%%%%%%%%%%
\vspace{20pt}

%%%%%%%%%%%%%%%%%%%%%%%%%%%%%%%%%%%%%%%%%%
\noindent \textbf{Author Contributions:} Conceptualization, D.F.; formal analysis, A.C.; writing-original draft preparation, A.C. and D.F.; writing-review and editing, M.C. and M.S.; supervision, M.S. All authors have read and agreed to the published version of the manuscript.

\vspace{6pt} 
%%%%%%%%%%%%%%%%%%%%%%%%%%%%%%%%%%%%%%%%%%
\noindent \textbf{Funding:} Authors would like to acknowledge the contribution of the European Union-NextGenerationEU through the "Quantum Busses for Coherent Energy Transfer'' (QUBERT) project, in the framework of Curiosity Driven 2021 initiative of the University of Genova.

\vspace{6pt}

\noindent \textbf{Acknowledgments:} Authors would like to thank G. Gemme for useful discussions.

%%%%%%%%%%%%%%%%%%%%%%%%%%%%%%%%%%%%%%%%%%
\vspace{6pt}
\noindent \textbf{Conflicts of Interest:} The authors declare no conflict of interest. 

%%%%%%%%%%%%%%%%%%%%%%%%%%%%%%%%%%%%%%%%%%
\vspace{6pt}\noindent \textbf{Abbreviations} \\
\noindent The following abbreviations are used in this manuscript:\\

\noindent 
\begin{tabular}{@{}ll}
QB & Quantum Battery\\
QHO & Quantum Harmonic Oscillator\\
TLS & Two-Level System\\
RWA & Rotating Wave Approximation
\end{tabular}

%%%%%%%%%%%%%%%%%%%%%%%%%%%%%%%%%%%%%%%%%%

%%%%%%%%%%%%%%%%%%%%%%%%%%%%%%%%%%%%%%%%%%


\begin{thebibliography}{999}
%
\bibitem{Riedel17} Riedel, M.F.; Binosi, D.; Thew, R.; Calarco, T. The European quantum technologies flagship programme. \emph{Quantum Sci. Technol.} \textbf{2017}, \emph{2}, 030501.
%
\bibitem{Acin18} Ac\'in, A.; Bloch, I.; Buhrman, H.; Calarco, T.; Eichler, C.; Eisert, J.; Esteve, D.; Gisin, N.; Glaser, S.J.; Jelezko, F.; et al. The quantum technologies roadmap: A European community view. \emph{New J. Phys.} \textbf{2018}, \emph{20}, 080201.
%
\bibitem{Raymer19} Raymer, M. G.; Monroe, C. The US quantum initiative. \emph{Quantum Sci. Technol.} \textbf{2019}, \emph{4}, 020504.
%
\bibitem{Esposito09} Esposito, M.; Harbola, U.; Mukamel, S. Nonequilibrium fluctuations, fluctuation theorems, and counting statistics in quantum systems. \emph{Rev. Mod. Phys.} \textbf{2009}, \emph{81}, 1665.
%
\bibitem{Vinjanampathy16} Vinjanampathy, S.; Anders, J. Quantum Thermodynamics. \emph{Contemp. Phys.} \textbf{2016}, \emph{57}, 545.
%
\bibitem{Campisi16} Campisi, M.; Fazio, R. Dissipation, correlation and lags in heat engines. \emph{J. Phys. A Math. Theor.} \textbf{2016}, \emph{49}, 345002.
%
\bibitem{Benenti17} Benenti, G.; Casati, G.; Saito, K.; Whitney, R.S. Fundamental aspects of steady-state conversion of heat to work at the nanoscale. \emph{Phys. Rep.} \textbf{2017}, \emph{694}, 1.
%
\bibitem{Campisi17} Campisi, M.; Goold, J. Thermodynamics information scrambling. \emph{Phys. Rev. E} \textbf{2017}, \emph{95}, 062127.
%
\bibitem{Campaioli18} Campaioli, F.; Pollock, F.A.; Vinjanampathy, S. \emph{Thermodynamics in the Quantum Regime};  Binder, F., Correa, L. A., Gogolin, C., Anders, J., Adesso, G., Eds.; Springer: Berlin/Heidelberg, Germany, 2018. 
%
\bibitem{Carrega22} Carrega, M.; Cangemi L. M; De Filippis, G.; Cataudella, V.; Benenti, G.; Sassetti, M. Engineering Dynamical Coupling for Quantum Thermodynamic Tasks. \emph{Phys. Rev. X Quantum} \textbf{2022}, \emph{3}, 010323.
%
\bibitem{Cavaliere22} Cavaliere, F.; Carrega, M.; De Filippis, G.; Cataudella, V.; Benenti, G.; Sassetti, M. Dynamical heat engines with non-Markovian reservoirs. \emph{
Phys. Rev. Res.} \textbf{2022} \emph{4}, 033233.

\bibitem{Alicki13} Alicki, R.; Fannes, M. Entanglement boost for extractable work from ensembles of quantum batteries. \emph{Phys. Rev. E} \textbf{2013}, \emph{87}, 042123.
%
\bibitem{Binder15} Binder, F.C.; Vinjanampathy, S.; Modi, K.; Goold, J. Quantacell: Powerful charging of quantum batteries. \emph{New J. Phys.} \textbf{2015}, \emph{17}, 075015.
%
\bibitem{Campaioli17} Campaioli, F.; Pollock, F.A.; Binder, F.C.; Celeri, L.; Goold, J.; Vinjanampathy, S.; Modi, K. Enhancing the Charging Power of Quantum Batteries. \emph{Phys. Rev. Lett.} \textbf{2017}, \emph{118}, 150601.
%
\bibitem{Friis18} Friis, N.; Huber, M. Precision and Work Fluctuations in Gaussian Battery Charging. \emph{Quantum} \textbf{2018}, \emph{2}, 61.
%
\bibitem{JF20} Juli\'a-Farr\'e, S.; Salamon, T.; Riera, A.; Bera, M.N.; Lewenstein, M. Bounds on the capacity and power of quantum batteries. \emph{Phys. Rev. Res.} \textbf{2020}, \emph{2}, 023113.
%
\bibitem{Le18} Le, T.P.; Levinsen, J.; Modi, K.; Parish, M.M.; Pollock, F.A. Spin-chain model of a many-body quantum battery. \emph{Phys. Rev. A} \textbf{2018}, \emph{97}, 022106.
%
\bibitem{Andolina18} Andolina, G. M.; Farina, D.; Mari, A.; Pellegrini, V.; Giovannetti, V.; Polini, M. Charger-mediated energy transfer in exactly solvable models for quantum batteries. \emph{Phys. Rev. B} \textbf{2018}, \emph{98}, 205423.
%
\bibitem{Rosa20} Rosa, D.; Rossini, D.; Andolina, G. M.; Polini, M., Carrega, M. Ultra-stable charging of fast-scrambling SYK quantum batteries. \emph{J. High Energy Phys.} \textbf{2020}, \emph{2020}, 67.
%
\bibitem{Crescente20} Crescente, A.; Carrega, M.; Sassetti, M.; Ferraro, D. Charging and energy fluctuations of a driven quantum battery. \emph{New J. Phys.} \textbf{2020}, \emph{22}, 063057.
%
\bibitem{Carrega20} Carrega, M.; Crescente, A.; Ferraro, D.; Sassetti, M. Dissipative dynamics of an open quantum battery. \emph{New J. Phys.} \textbf{2020}, \emph{22},~083085.
%
\bibitem{Mitchison21} Mitchison, M.T.; Goold, J.; Prior, J. Charging a quantum battery with linear feedback control. \emph{Quantum} \textbf{2021}, \emph{5},~500.
%
\bibitem{Caravelli21} Caravelli, F.; Yan, B.; Garc\'ia-Pintos, L.P.; Hamma, A. Energy storage and coherence in closed and open quantum batteries. \emph{Quantum} \textbf{2021}, \emph{5}, 505.
%
\bibitem{Seah21} Seah, S.; Perarnau-Llobet, M.; Haack, G.; Brunner, N.; Nimmrichter S. Quantum Speed-Up in Collisional Battery Charging. \emph{Phys. Rev. Lett.} \textbf{2021}, \emph{127}, 100601.
%
\bibitem{Gyhm22} Gyhm, J.-Y.; \v{S}afr\'anek, D.; Rosa, D. Quantum Charging Advantage Cannot Be Extensive without Global Operations. \emph{Phys. Rev. Lett.} \textbf{2022}, \emph{128}, 140501.
%
\bibitem{Crescente22} Crescente, A.; Ferraro, D.; Carrega, M.; Sassetti, M. Enhancing coherent energy transfer between quantum devices via a mediator. \emph{Phys. Rev. Res.} \textbf{2022}, \emph{4}, 033216.
%
\bibitem{Shaghaghi23} Shaghaghi, V.; Singh, V.; Carrega, M.; Rosa, D.; Benenti, G. Lossy Micromaser Battery: Almost Pure States in the Jaynes-Cummings Regime. \emph{Entropy} \textbf{2023}, \emph{25}, 430.
%
\bibitem{Santos23} Santos, T.F.F.; de Almeida, Y.V.; Santos, M.F. Vacuum-enhanced charging of a quantum battery. \emph{Phys. Rev. A} \textbf{2023}, \emph{107}, 032203.
%
\bibitem{Ferraro18} Ferraro, D.; Campisi, M.; Andolina, G. M.; Pellegrini, V.; Polini, M. High-Power Collective Charging of a Solid-State Quantum Battery. \emph{Phys. Rev. Lett.} \textbf{2018}, \emph{120}, 117702.
%
\bibitem{Ferraro19} Ferraro, D.; Andolina, G.M.; Campisi, M.; Pellegrini, V.; Polini, M.; Quantum supercapacitors. \emph{Phys. Rev. B} \textbf{2019}, \emph{100}, 075433.
%
\bibitem{Andolina19} Andolina, G. M.; Keck, M.; Mari, A.; Campisi, M.; Giovannetti, V.; Polini, M. Extractable Work, the Role of Correlations, and Asymptotic Freedom in Quantum Batteries. \emph{Phys. Rev. Lett.} \textbf{2019}, \emph{122}, 047702.
%
\bibitem{Crescente20b} Crescente, A.; Carrega, M.; Sassetti, M.; Ferraro, D. Ultrafast charging in a two-photon Dicke quantum battery. \emph{Phys. Rev. B} \textbf{2020}, \emph{102}, 245407.
%
\bibitem{Delmonte21} Delmonte, A.; Crescente, A.; Carrega, M.; Ferraro, D.; Sassetti, M. Characterization of a two-photon quantum battery: Initial conditions, stability and work extraction. \emph{Entropy} \textbf{2021}, \emph{23}, 612.
%
\bibitem{Erdman22} Erdman, P.A.; Andolina G.M.; Giovannetti, V.; No\'e F. Reinforcement learning optimization of the charging of a Dicke quantum battery. \emph{arXiv} \textbf{2022}, \emph{arXiv:2212.12397}.
%
\bibitem{Dou23} Dou, F.-Q.; Yang, F.-M. Superconducting transmon qubit-resonator quantum battery. \emph{Phys. Rev. A} \textbf{2023}, \emph{107}, 023725.
%
\bibitem{Gemme23} Gemme, G.; Andolina, G.M.; Pellegrino, F.M.D.; Sassetti, M.; Ferraro, D. Off-resonant Dicke Quantum Battery: Charging by Virtual Photons. \emph{Batteries} \textbf{2023}, \emph{9}, 197. 
%
\bibitem{Quach22} Quach, J.Q.; McGhee, K.E.; Ganzer, L.; Rouse, D.M.; Lovett, B.W.; Gauger, E.M.; Keeling, J.; Cerullo, G.; Lidzey, D.G.; Virgili, T. Superabsorption in an organic microcavity: Toward a quantum battery. \emph{Sci. Adv.} \textbf{2022}, \emph{8}, eabk3160.
%
\bibitem{Hu22} Hu, C.-K.; Qiu, J.; Souza, P.J.P.; Yuan, J.; Zhou, Y.; Zhang, L.; Chu, J.; Pan, X.; Hu, L.; Li, J.; et al. Optimal charging of a superconducting quantum battery. \emph{Quantum Sci. Technol.} \textbf{2022}, \emph{7}, 045018.
%
\bibitem{Gemme22} Gemme, G.; Grossi, M.; Ferraro, D.; Vallecorsa, S.; Sassetti, M. IBM quantum platforms: A quantum battery perspective. \emph{Batteries} \textbf{2022}, \emph{8}, 43.
%
\bibitem{Scarlino19} Scarlino, P.; van Woerkom, D.J.; Mendes, U.C.; Koski, J.V.; Landig, A.J.; Andersen, C.K.; Gasparinetti, S.; Reichl, C.; Wegscheider, W.; Ensslin, K.; et al. Coherent microwave-photon-mediated coupling between a semiconductor and a superconducting qubit. \emph{Nat. Commun.} \textbf{2019}, \emph{10}, 3011.
%
\bibitem{Landig19} Landig, A.J.; Koski, J.V.; Scarlino, P.; M\"oller, C.; Abadillo-Uriel, J.C.; Kratochwil, B.; Reichl, C.; Wegscheider, W.; Coppersmith, S.N.; Friesen, M.; et al. Virtual-photon-mediated spin-qubit transmon coupling. \emph{Nat. Commun.} \textbf{2019}, \emph{10}, 5037.
%
\bibitem{Arrachea23} Arrachea, L. Energy dynamics, heat production and heat-work conversion with qubits: Towards the development of quantum machines. \emph{Rep. Prog. Phys.} \textbf{2023}, \emph{86}, 036501.
%
\bibitem{Castro08} Olaya-Castro, A.; Lee, C.F.; Fassioli Olsen, F.; Johnson, N.F. Efficiency of energy transfer in a light-harvesting system under quantum coherence. \emph{Phys. Rev. B} \textbf{2008}, \emph{78}, 085115.
%
\bibitem{Sahoo11} Sahoo, H. F\"orster resonance energy transfer - A spectroscopic nanoruler: Principle and applications. \emph{J. Photochem. Photobiol. C Photochem. Rev.} \textbf{2011}, \emph{12}, 1.
%
\bibitem{Auffeves22} Auff\'eves, A. Quantum Technologies Need a Quantum Energy Initiative. \emph{PRX Quantum} \textbf{2022}, \emph{3}, 020101.
%
\bibitem{Stevens22} Stevens, J.; Szombati, D.; Maffei, M.; Elouard, C.; Assouly, R.; Cottet, N.; Dassonneville, R.; Ficheux, Q.; Zeppetzauer, S.; Bienfait, A.; et al. Energetics of a Single Qubit Gate. \emph{Phys. Rev. Lett.} \textbf{2022}, \emph{129}, 110601.
%
\bibitem{Lewis23} Lewis, D.; Moutinho, J. P.; Costa, A.; Omar, Y.; Bose, S. Low-Dissipation Data Bus via Coherent Quantum Dynamics. \emph{arXiv}~\textbf{2023}, {arXiv:2304.02391v1}.
\bibitem{Laucht21} Laucht, A.; Hohls, F.; Ubbelohde, N.; Gonzalez-Zalba, M.F.; Reilly, D.J.; Stobbe, S.; Schroder, T.; Scarlino, P.; Koski, J.V.; Dzurak, A. Roadmap on quantum nanotechnologies. \emph{Nanotechnology} \textbf{2021}, \emph{32}, 162003.
%
\bibitem{Sung21} Sung, Y.; Ding, L.; Braum\"uller, J.; Veps\"al\"inen, A.; Kannan, B.; Kjaergaard, M.; Greene, A.; Samach, G.O.; McNally, C.; Kim, D.; et al. Realization of High-Fidelity CZ and ZZ-Free iSWAP Gates with a Tunable Coupler. \emph{Phys. Rev. X} \textbf{2021}, \emph{11}, 021058. 
%
\bibitem{Makhlin01} Makhlin, Y.; Sch\"on, G.; Shnirman, A. Quantum-state engineering with Josephson-junction devices. \emph{Rev. Mod. Phys.} \textbf{2001}, \emph{73}, 357.
%
\bibitem{Weiss} Weiss, U. \emph{Quantum Dissipative Systems,} 4th ed.; World Scientific: Singapore, 2012.
%
\bibitem{Grifoni95} Grifoni, M.; Sassetti, M.; Hanggi, P.; Weiss, U. Cooperative effects in the nonlinearly driven spin-boson system. \emph{Phys. Rev. E} \textbf{1995}, \emph{52}, 3596.
%
\bibitem{Calzona23} Calzona, A.; Carrega, M. Multi-mode architectures for noise-resilient superconducting qubits. \emph{Supercond. Sci. Technol.} \textbf{2023}, \emph{36}, 023001.
%
\bibitem{Devoret13} Devoret, M.H.; Schoelkopf, R.J. Superconducting Circuits for Quantum Information: An Outlook. \emph{Science} \textbf{2013}, \emph{339}, 1169.
%
\bibitem{Wendin17} Wendin, G. Quantum information processing with superconducting circuits: A review. \emph{Rep. Prog. Phys.} \textbf{2017}, \emph{80}, 106001.
%
\bibitem{Krantz19} Krantz, P.; Kjaergaard, M.; Yan, F.; Orlando, T.P.; Gustavsson, S.; Oliver, W. D.; A quantum engineer's guide to superconducting qubits. \emph{Appl. Phys. Rev.} \textbf{2019}, \emph{6}, 021318.
%
\bibitem{Schweber67} Schweber, S. On the application of Bergmann Hilbert spaces to dynamical problems. \emph{Ann. Phys.} \textbf{1967}, \emph{41}, 205.
%
\bibitem{Graham84} Graham, R.; H\"ohnerbach, M. Two-state system coupled to a boson mode: Quantum dynamics and classical approximations. \emph{Z. Phys. B} \textbf{1984}, \emph{57}, 233.
%
\bibitem{Schleich} Schleich, W.P. \emph{Quantum Optics in Phase Space}; Wiley-VCH: Berlin, Germany, 2018.
%
\bibitem{Thomas22} Thomas, G.; Pekola, J.P. Dynamical phase and quantum heat transport at fractional frequencies. \emph{arXiv} \textbf{2022}, {arXiv:2207.07632}. 
%
\bibitem{Alexander20} Alexander, T.; Kanazawa, N.; Egger, D.J.; Capelluto, L.; Wood, C.J.; Javadi-Abhari, A.; McKay, D.C. Qiskit pulse: Programming quantum computers through the cloud with pulses. \emph{Quantum Sci. Technol.} \textbf{2020}, \textit{5}, 044006.


















\end{thebibliography}
\end{document}